\documentclass[sigconf]{acmart}

\usepackage{times}
\usepackage{helvet}
\usepackage{courier}
\usepackage{subcaption}
\usepackage[font={small}]{caption}
\usepackage{graphicx}
\usepackage[toc,page,header]{appendix}
\usepackage{minitoc}
\usepackage{comment}

\AtBeginDocument{%
  \providecommand\BibTeX{{%
    \normalfont B\kern-0.5em{\scshape i\kern-0.25em b}\kern-0.8em\TeX}}}

\setcopyright{none}
\renewcommand\footnotetextcopyrightpermission[1]{}
\acmConference[AIES '21] {Proceedings of the 2021 AAAI/ACM Conference on AI, Ethics, and Society}{May 19--21, 2021}{Virtual Event, USA.}

\begin{document}

\title[On the Validity of Arrest as a Proxy for Offense]{On the Validity of Arrest as a Proxy for Offense:\\Race and the Likelihood of Arrest for Violent Crimes}

\author{Riccardo Fogliato}
\email{rfogliat@andrew.cmu.edu}
\affiliation{%
  \institution{Carnegie Mellon University}
  \streetaddress{}
  \city{}
  \state{}
  \country{}}

\author{Alice Xiang}
\affiliation{%
  \institution{Sony AI}
  \streetaddress{}
  \city{}
  \country{}}
\email{}

\author{Zachary Lipton}
\affiliation{%
  \institution{Carnegie Mellon University}
  \streetaddress{}
  \city{}
  \state{}
  \country{}}

\author{Daniel Nagin}
\affiliation{%
  \institution{Carnegie Mellon University}
  \streetaddress{}
  \city{}
  \state{}
  \country{}}

\author{Alexandra Chouldechova}
\affiliation{%
  \institution{Carnegie Mellon University}
  \streetaddress{}
  \city{}
  \state{}
  \country{}}
  
\newcommand\blfootnote[1]{%
  \begingroup
  \renewcommand\thefootnote{}\footnote{#1}%
  \addtocounter{footnote}{-1}%
  \endgroup
}

\renewcommand{\shortauthors}{Fogliato, Xiang, Lipton, Nagin, Chouldechova}


\begin{abstract}
The risk of re-offense is considered in decision-making at many stages of the criminal justice system, from pre-trial, to sentencing, to parole.  To aid decision makers in their assessments, institutions increasingly rely on algorithmic risk assessment instruments (RAIs). These tools assess the likelihood that an individual will be arrested for a new criminal offense within some time window following their release. However, since not all crimes result in arrest, RAIs do not directly assess the risk of re-offense.  Furthermore, disparities in the likelihood of arrest can potentially lead to biases in the resulting risk scores. Several recent validations of RAIs have therefore focused on arrests for violent offenses, which are viewed as being more accurate reflections of offending behavior. In this paper, we investigate biases in violent arrest data by analysing racial disparities in the likelihood of arrest for White and Black violent offenders. We focus our study on 2007--2016 incident-level data of violent offenses from 16 US states as recorded in the National Incident Based Reporting System (NIBRS). Our analysis shows that the magnitude and direction of the racial disparities depend on various characteristics of the crimes. In addition, our investigation reveals large variations in arrest rates across geographical locations and offense types. We discuss the implications of the observed disconnect between re-arrest and re-offense in the context of RAIs and the challenges around the use of data from NIBRS to correct for the sampling bias.

\end{abstract}


\maketitle

\section{Introduction}\label{sec:intro}

\blfootnote{Accepted at AAAI/ACM Conference on Artificial Intelligence, Ethics, and Society (AIES), 2021.}

Recidivism risk assessment instruments (RAIs) are increasingly used to inform decisions throughout the criminal justice system \citep{metz2020algorithm}. To justify their adoption,
practitioners, vendors, and scholars 
often claim that the use of RAIs can lead to more objective, transparent, and fair
decisions \citep{kleinberg2018human, goel2018accuracy, mullainathan2019biased}.
However, the ability of RAIs to help achieve these ambitious goals
has been contested \citep{barabas2018interventions, green2020false}.  A major concern is that the RAIs themselves may exhibit problematic predictive biases.  In May
2016, an investigation by ProPublica examined COMPAS, an RAI used in Broward
County, Florida, finding that the tool's predictions exhibited a higher false
positive rate for Black (vs. White) defendants and thus concluding that COMPAS
was biased against Black defendants~\citep{propublica2016}. Critics rebutted
these claims, arguing that error rate imbalance did not necessarily indicate
racial bias and that RAIs should be assessed for predictive parity and
calibration, two properties that COMPAS satisfied
\citep{floresfalse,dieterich2016compas}. Later work on the topic showed that
predictive parity and error rate balance cannot be simultaneously satisfied when
the recidivism rates differ across demographic groups
\citep{chouldechova2017fair,kleinberg2016inherent,corbett2017algorithmic},
sparking a flurry of technical research on algorithmic fairness within the
machine learning community. These papers, many of which draw on the ProPublica-COMPAS
dataset, tend to focus on metrics for capturing bias, techniques for mitigating
bias as measured by these metrics, and characterizing fundamental tradeoffs
\citep{hardt2016equality, berk2017convex, donini2018empirical,
agarwal2018reductions, johndrow2019algorithm, celis2019classification}.

Problematically, much of this work rests on the assumption that the observed outcomes (a.k.a. dependent variables)
represent ground truth.   Potential sources of bias in the training data are often overlooked.
For example, many bias mitigation strategies amount to equalizing the RAI's
performance across racial groups with respect to some metric based on
\textit{re-arrest} outcomes. However, it is widely acknowledged that arrest data are
affected by {\it sampling bias}\footnote{Sampling bias arises when not all
elements of the population of interest are equally likely to be sampled. In the
presence of sampling bias, the characteristics of the collected sample of
observations, which is said to be affected by ``sample bias'', are not
representative of the characteristics of the population from which the sample is
drawn.}~\cite{brame2004criminal} and re-arrest may represent an imperfect
proxy for the target of interest, \textit{re-offense}. Importantly, there is ample
evidence that racial minorities tend to face higher risk of
arrest, especially for crimes targeted through proactive policing, such as drug
and traffic offenses \citep{lum2016predict,goel2016precinct,pierson2020large}.
Thus, individuals with the same probability of re-offense may nevertheless have
different probabilities of re-arrest.
RAIs trained on such data may appear to be fair predictors of re-arrest but nevertheless be unfair, even by the same metrics, were they
to be assessed on re-offense~\citep{fogliato20a}.

In recognition of this potential discrepancy, recent work on model validation has primarily examined the predictive bias of RAIs as predictors of re-arrest for \textit{violent} crimes.  Such arrests are viewed as representing ``the most unbiased criterion [of offense] available''~\citep{skeem2016risk}. \citet{beck2018racial}, for instance, concluded that statistics based on arrests for violent offenses could be an (almost) racially-unbiased proxy of criminal involvement. These arguments are
generally premised 
upon three central pieces of evidence \citep{walsh2004race}. 
First, the racial composition of violent offenders recorded in victimization reports roughly matches that reported in arrest data 
\citep{hindelang1978race,tonry1995malign,beck2018racial}. 
Second, several analyses of violent offenses known to law enforcement have
found that Black offenders are arrested at \textit{lower} rates than White offenders \citep{d2003race, pope2003race}. 
Third, unlike in the case of lower level crimes, in the case of violent crimes police are granted much more limited discretion on whether to make an arrest.

This body of evidence suggests that, overall, White violent offenders are at least as likely as Black violent offenders to be arrested, and hence that the
overrepresentation of Blacks among arrestees is attributable to differential involvement in
offending. The cited evidence does not, however, indicate that all offenses are
equally likely to result in arrests, i.e., that, conditional on situational and
contextual factors, the likelihood of arrest is equal across racial groups. 
In absence of such evidence, it remains unclear whether re-arrest outcomes could
be considered a reliable proxy for re-offense. 

In this paper, we show that the likelihood of arrest for violent crimes varies
with the characteristics of the offense, including the offender's race. Our
analysis is based on incident-level data of violent offenses involving lone
victims and offenders as reported by police
agencies to the FBI in 16 US states between 2007 and 2016 through the National
Incident Based Reporting System (NIBRS), a national crime data collection
program.  We focus on incidents involving only Black or White victims and offenders.  

We find that arrest rates vary substantially across offense types and
geographical locations. Controlling for the state, White and Black offenders are
arrested at similar rates for crimes of forcible rape and murder. For assaults
and robbery, however, arrest rates are generally higher for White offenders.
This counterintuitive pattern is largely explained by the lower arrest rates in
jurisdictions with larger shares of Black offenders. We offer two potential
explanations for these findings. First, consistent with the hypothesis of {\it
benign neglect} and based on a perspective of victim devaluing, fewer policing
resources are allocated to solving the least serious forms of violent crimes
that involve Black victims. Second, due to higher levels of {\it legal
cynicism}, which can be traced back to a history of over-policing and systemic
discrimination, Black communities may be less likely to cooperate with law
enforcement. Through a regression analysis, we assess whether the observed racial
disparities in the likelihood of arrest can be explained by crimes
characteristics. Due to the (perhaps unavoidable) misspecification of our
modeling approach, the strength and direction of the association between the
offender's race and the likelihood of arrest depend on the sample of crimes that
is considered. These findings indicate that simple models targeting the
``effect'' of race, which have been used in a plethora of studies like ours, may
not fully capture the complexities of the role of race in arrests.

The large variations in the likelihood of arrest across crimes characteristics,
namely geographical areas and offense types, together with the presence of
racial disparities, call into question the reliability of arrests as a proxy for
violent offending. We discuss how the sampling biases observed in the data can
affect the training and assessment of RAIs, leading to the severe
underestimation of the probability of re-offense for certain subgroups of
offenders (e.g., sex offenders). Finally, we discuss why the observed sampling bias cannot
be corrected using data from NIBRS alone.

\section{Background}\label{sec:related-work}

Prior research studying racial disparities in the likelihood of arrest for violent
offenses has generally fallen into two broad categories: (i) macro-level analyses studying the
relationship between arrest rates and geographical or socioeconomic factors;
and (ii) micro-level analyses examining associations between between the likelihood of arrest and
the characteristics of the incident and parties involved.

Macro-level studies of the
likelihood of arrest tend to fall into one of two (not mutually exclusive)
theoretical frameworks: minority threat and benign neglect.   The {\it minority threat}
hypothesis~\cite{blalock1967toward,kirk2011legal} asserts that the
majority group (Whites) attempts to control the minority
group (e.g., Blacks) by imposing stronger formal mechanisms of social control,
such as policing. According to this theory, we might expect to observe a positive
association between the share of Black residents in an area and the size of the police force and arrest rates. The correlation between these factors
may be weaker when interactions between the majority and the minority are
limited, such as in regions where racial segregation is strongest
\citep{stults2007racial}. 
The \emph{benign neglect} hypothesis
\citep{liska1984social} posits that, as the size of the Black population grows,
rates of intraracial crime are likely to increase as well. Whites may then weaken the
mechanisms of social control on the Black population in racially segregated neighborhoods
through a reduction in policing resources. As a result, Black victims would face
difficulties in legitimating their complaints and convincing police officers to
take action~\cite{wilson1978varieties}. Under this hypothesis, we might expect
arrest rates to be lower in predominantly Black neighborhoods. 
The two
frameworks emphasize different socioeconomic drivers of differential arrest: While the minority threat
hypothesis focuses on the role of race as a threat (be it economical or
political), the benign neglect hypothesis emphasizes the devaluation of Black
victims. Both hypotheses have received mixed empirical support
\cite{ousey2008racial}.
The size of the police force has been shown to be positively associated with the share of Black population in the area \cite{kent2005minority, stults2007racial}. \citet{kent2005minority} found that the most highly racially segregated cities with larger Black populations tend to have smaller police forces, whereas \citet{stults2007racial} reported a positive association between the level of racial segregation and the size of police forces.
Other studies have reported that arrest rates for Black offenders are
negatively associated with the share of Black residents in the city
\citep{parker2005racial}, negatively associated with the level of racial
segregation \citep{liska1985testing,stolzenberg2004multilevel}, and positively
associated with the share of interracial crime \citep{eitle2002racial}. Arrest
rates for homicide have been shown to be lower in predominantly Black
neighborhoods \citep{puckett2003factors,fagan2018police}, and higher in large
cities with substantial socioeconomic disparities between the two groups
\citep{borg2001mobilizing}. In our analysis, we also found that arrest rates are
lower in police agencies with larger proportions of Black offenders.

The second area of work focuses on racial disparities at the micro level.
Here, researchers typically attempt to isolate the effects of police
discrimination from other sources of disparity by controlling for legal
non-discretionary  factors through regression analysis. The variability in
the likelihood of arrest that is explained by features such as victim and offender's race is then interpreted as (potentially) resulting from
discriminatory practices.  These analyses are inherently limited by the extent to which available data capture all alternative explanatory factors, and the challenge of interpreting race as a causal or mediating factor in regression \citep{vanderweele2014causal}. 
Research in this area has
primarily relied on three sources of data: (i) field observations  of
encounters between police officers and citizens; (ii) self-reports of offending
behavior; (iii) and official crime incidents records reported by police agencies.  We focus here on (iii), as it is most closely related to our work.  Prior studies of single victim, single offender incidents in the NIBRS data found that robberies, simple assault,
aggravated assaults were more likely to be cleared by arrest when the offender is White than when the offender is Black, even after adjusting for many contextual and situational factors of the incident
\citep{d2003race,pope2003race,roberts2009victim}. These three types of offenses
constitute the majority of violent crimes reported to police. This association did not persist for the most serious violent crimes of forcible rape and murder/non-negligent manslaughter.  In their analysis of rape offenses, \citet{d2003race} found that the overall clearance rate for Black
offenders was marginally higher than for Whites, but that there was no statistically significant difference after controlling for
situational factors. Through survival analysis modeling, \citet{roberts2009victim} found that murder/non-negligent manslaughter incidents with
non-White offenders were more likely to result in an arrest. 
Lastly, in a recent analysis of violent offenses of 2003--2012 NIBRS data
including crimes committed by multiple offenders, \citet{lantz2019co} found that incidents were less likely to result in arrests when the offender was Black. However, when they restricted to
incidents where Black and White individuals offended together, they found that
Black offenders were slightly more likely to be arrested. Based on this
evidence, the authors argued that omitted variable bias might affect regression
analyses conducted on NIBRS data. 
Our study of violent offenses on NIBRS is
closely related to that of \citet{d2003race}, but our analysis is much larger scale, includes incidents of murder/non-negligent manslaughter, and, unlike
theirs, does not rely on the (unrealistic) assumption that the regression model
is well-specified. Unlike \citet{lantz2019co}, we focus specifically
on geographical variations in the likelihood of arrest and on the issue of model
misspecification.  We find that the sign of the offender race coefficient varies with the subset of crimes used in the analysis, which suggests that the conclusions of prior work are dependent on model specification. 
\section{Data}
Our analysis is mainly based on incident-level data of offenses from the
National Incidents Based Reporting System (NIBRS) and data of police agencies
from the Law Enforcement Employees Report. We describe each of the two datasets
in turn. 

\paragraph{NIBRS data}
NIBRS is part of the FBI's Uniform Crime Reporting (UCR) program. Through NIBRS,
law enforcement agencies submit detailed data on the characteristics of
incidents that are known to them, such as the demographics of victims and
offenders \cite{nibrs_manual19}. In our
analysis, we use incident-level data of offenses and arrests recorded in NIBRS
between 2007 and 2016 by aggregating annual data files obtained from the
Inter-University Consortium for Political and Social Research (ICPSR)
\citep{nibrs07,nibrs08,nibrs09,nibrs10,nibrs11,nibrs12,nibrs13,nibrs14,nibrs15,nibrs16}.
We only consider incidents that include at least one offense of
murder/non-negligent manslaughter, forcible rape, robbery, aggravated assault,
or simple assault. 

In order to conduct our statistical analysis, we process the data as follows.
First, we keep only data from the 16 states that submitted all their crime data
through NIBRS in 2014: Arkansas (AR), Colorado (CO), Delaware (DE), Idaho (ID),
Iowa (IA), Kentucky (KY), Michigan (MI), Montana (MT), New Hampshire (NH), North
Dakota (ND), South Carolina (SC), South Dakota (SD), Tennessee (TN), Vermont
(VT), Virginia (VA), and West Virginia (WV). Our results may not generalize beyond these 16 states. 
%
Second, following the approach of~\citet{d2003race}, we keep only incidents that
involve only one offender and one victim. The exclusion of incidents with
multiple victims or offenders is mainly motivated by the additional assumptions
that would be required if these types of incidents were considered for the
analysis. Since data from NIBRS do not contain offender-level characteristics on
the incident (e.g., which offender used the weapon), an analysis of
multi-offender crimes would need to rely on assumptions around the
circumstances of the incidents and require more complex regression models, e.g.,
the dependence of observations corresponding to the same incident should be modeled. 
By excluding these incidents, the analysis of the interactions between victims'
and offenders' relevant features is simplified.
Importantly, we consider only incidents where both the victim's and offender's
races are recorded as either Black or White. We cannot exclude Hispanics from
the analysis because the ethnicity field was introduced in the data only in 2012
and is often left empty. We note that almost every victim of Hispanic ethnicity
is recorded as being White. 
Lastly, we drop all incidents that are cleared by exceptional means. We discuss
this type of clearance in Appendix \ref{sec:additional_data}.
The final dataset includes 9,181 incidents of murder/non-negligent manslaughter,
103,309 forcible rapes, 101,133 robberies, 596,324 aggravated assaults, and
2,669,399 simple assaults.

For the regression analysis, we transform the features (a.k.a. covariates, regressors, independent variables) as follows. 
We first code dummies corresponding to the age, sex, and race of victim and offender.
We create victim-level dummy variables to indicate whether
the victim suffered a serious or a minor injury, knew the offender (i.e., the
offender was not unknown to victim or was a stranger), and whether the incident
occurred in the residence of the victim. 
We also create binary variables for incident's characteristics to indicate whether
the incident occurred during the day (i.e., between 7am and 8pm), a firearm was
involved, a weapon other than a firearm was involved, the offense was only
attempted but not completed (only available for forcible rape and robbery), the offender was suspected of having used drugs or alcohol, and whether
ancillary offenses were committed. Finally, we create a variable that corresponds to the
share of Black violent offenders (out of all violent offenders) for each police
agency and a series of dummy variables for states and years to capture state and year fixed effects.

\paragraph{Law enforcement data}  We match data from NIBRS with information
regarding the police agency from the Law Enforcement Employees Report. This data are obtained from ICSPR for the years 2007--2016~\cite{leoka07,leoka08,leoka09,leoka10,leoka11,leoka12,leoka13,leoka14,leoka15,leoka16}. For each police agency we extract information on the total
population served and the number of police officers employed.

\section{Methodology}\label{sec:methods}

Throughout the analysis, we assume the
observations (i.e., the incidents) to be independent. Although each observation
corresponds to a single crime incident and we exclude incidents that are cleared
by exceptional means, independence may be violated if multiple
incidents involved the same offender. Since offenders' identifiers are not
available in the data, we are unable to model this potential dependence.

In order to test for macro-level variations in racial disparities across
geography, we split our data by jurisdiction (i.e., police agency) and examine
the arrest rates for different types of crimes and demographics. In our
analysis, the arrest rate corresponds to the share of offenses that result in
arrests. For example, the arrest rate for White offenders is given by the ratio
between the number of arrests and the number of offenses involving White
offenders. Such summary statistics likely represent upper bounds for the arrest
rates in the entire population of crimes, i.e., also the crimes that are not
accounted for by our analysis such as those that are not reported to law enforcement. We assess the correlation between variables via
Pearson's correlation coefficient ($\rho$). Statistical significance is assessed at the $0.01$ level for all hypothesis tests.

Since not all of the data features are fully observed, for our regression analysis we need to carry out data imputation.  
The datasets (one for each offense type) contain small shares of missing values consisting of up to approximately 5\% of all observations, with the exception of around 20\% in case of the offender’s age in offenses of robbery. As in past work~\cite{d2003race}, we
assume the data to be missing at random (MAR)~\cite{rubin1976inference}. 
We impute ten datasets via multiple imputation by chained
equations~\cite{azur2011multiple}.  
On each imputed dataset, we use logistic regression to model
the dependence of the likelihood of arrest on race and other factors. We calculate sandwich standard errors for the coefficients estimates. We then obtain a single set of coefficients estimates and the corresponding
standard errors using the formulas provided by~\citet[pp.
76-77]{rubin2004multiple}.  This analysis is conducted separately for each offense type.  

In the second stage of our analysis, we investigate whether the logistic regression model appears to be well-specified. Here, we apply the approach proposed by \citet{buja2019models}, which is based on the principle that coefficients estimates of a correctly specified model do not change significantly as the distribution of the regressors varies. Changes in the
coefficients estimates under observation reweighting are therefore indicative of model misspecification. In our analysis, we employ the \textit{focal slope} visual model diagnostic introduced by \citet{buja2019models} to
assess how the offender's race coefficient estimate varies with
the reweighting of the distribution of certain regressors. This tool provides
insights into the interactions between the race variable and other regressors
without modeling the interactions directly in the regression model. Our aim is
to investigate whether the estimate of the coefficient relative to the
offender's race always has a consistent sign and magnitude under different reweightings. 

To implement the reweighting procedure, we proceed as follows. We first
construct a grid of ten evenly
spaced values for the numeric
features, and use the grid values of $\{0,1\}$ for the binary features. For each feature, we split the observations into groups based on the grid's cell center that is closest to each observation's feature value in absolute distance.  For each feature-grid cell pair, we then obtain 100 estimates of the logistic regression coefficients by bootstrap resampling observations from the given group. Due to the high computational cost driven by the large number of observations of simple assault incidents, we apply the focal slope diagnostics to only one of the ten imputed datasets for this offense type.  The coefficients
estimates on the ten imputed datasets are nearly identical, so we do not expect the focal slope analyses to vary with the imputed data sample. 
Our data analysis is fully reproducible. The full \texttt{R} \cite{team2013r} code
is available at the following address: \url{https://github.com/ricfog/on-the-validity-of-arrest}. 

\section{Limitations}\label{sec:limitations} Despite the wealth of information contained therein,
analyses of NIBRS data suffer from several key limitations. First, not every police agency
reports all of its crime data through NIBRS \citep{bibel2015considerations}. These omissions
may reflect partial reporting (e.g., for some period of time each year) or
complete non-reporting by agencies in the states considered. If such data
omissions occurred at random (with respect to crimes characteristics and outcomes), then our inference at the level of the police
agency would not be affected, but the representativeness of our sample with
respect to the population of interest (i.e., all the crime incidents known to
police agencies in the 16 states considered) could be impacted. Data
omissions that do not occur at random would be even more problematic. For example, the existence of different reporting protocols across states or jurisdictions, such as the non-reporting of incidents that do not result in arrest by certain agencies, could explain the large variations in arrest rates that we observe.
Importantly, there is evidence that cases that are considered as
``unfounded'' are not tracked by NIBRS~\cite{propublica2018} and consequently will not be included in the data. 
Another issue pertains to the quality of the data that are reported, which may
not offer an unbiased picture of the circumstances of the incident. Victims may
report mistaken or even false details regarding the incident. Likewise, data
from police departments may also be the artefact of a selection, manipulation,
and review process, as the study of \citet{richardson2019dirty} has shown. While
in many instances the arrestee is the offender, there may be incidents in the
data that are cleared by wrongful arrests. Due to the absence of data on which incidents might
correspond to wrongful arrests, we are unable to account for this possibility in our analysis. This may be problematic given the evidence that, even for the
most serious types of offenses, wrongful arrests may represent a considerable share of
all arrests~\cite{loeffler2019measuring}. 
Our analysis considers
only violent crime incidents that are not cleared by exceptional means and where there is a lone offender and victim whose race is reported as either
Black or White. Some of
the conclusions may therefore fail to generalise to other types
of incidents. 
It is important to keep in mind that the characteristics
of the sample of offenses in NIBRS is not representative of all crimes at
the national level~\cite{mccormack2017assessing, pattavina2017assessing}.
Lastly, we could not exclude Hispanics from our sample, nor could we code this group separately for our analysis. There is
evidence that the degree of involvement in violent offending of Hispanics falls
within the levels of the White and of the Black
populations \cite{steffensmeier2011reassessing}. As previously
mentioned, in NIBRS data Hispanics are for the most part included within the
White population. Any existing differences in offending behavior or arrest rates
between Hispanic and non-Hispanic White offenders will not be captured by our analysis. 
\section{Results}\label{sec:results}

Since, to the best of our knowledge, this study presents the first analysis of racial disparities in violent offenses using the most recent years of NIBRS data, we begin by providing some
summary statistics on arrest rates across victim and offender's demographics and
crime types (see also Table~\ref{tab:summary_stats_data}). 
Overall, 41\% of all offenders and 34\% of all victims are Black. As a reminder,
only Black and White individuals are represented in the dataset. The
overrepresentation of Black individuals among offenders is lowest in case of
incidents of forcible rape (31\%) and highest in case of robbery (76\%). 
We then examine the likelihood of arrest across offenders' racial groups. Arrests are more frequent in case of
incidents involving White offenders across all types of crimes (overall, 56\%
for White vs. 42\% for Black offenders). However, there exist differences in the gap in
arrest rates across types of offenses. The gap is largest in case of robbery
(36\% vs 19\%) and aggravated assault (62\% vs 44\%), followed by simple assault
(57\% vs 43\%), which constitute the majority of the offenses present in the
data. In contrast, the gap is small for murder/non-negligent manslaughter (73\%
vs 67\%) and rape (27\% vs 25\%). Lastly, we assess the racial
composition of victims. Intraracial crimes constitute four-fifths or more of
the offenses across all types of crimes, with the exception of rape (63\%) and
robbery (57\%) offenses committed by Black offenders. Interestingly, arrest
rates for interracial crimes are similar across racial groups (within 2\%) for all types of crimes other than robbery.
In summary, we observe that arrest rates are highest for intraracial offenses
among Whites, followed by interracial offenses, and last by intraracial offenses
among Black individuals.

\subsection{Macro-level variations in arrest rates} 

It is possible that the observed gap in arrest rates across offenders' racial
groups could be explained by variations in the likelihood of arrest and in the
racial composition of offenders across geographical areas, e.g., as in case of a
Simpson's paradox. To analyse such variation, we first consider the
data regarding simple assaults. The large size of this dataset allows us to
consider arrest rates at the level of the cities and police agencies. For this analysis, 
we focus on police agencies that reported at least 100 crimes of simple assault
in the ten-year period spanned by our data.

\begin{figure}[t] 
   \centering
      \includegraphics[width=\linewidth, keepaspectratio]{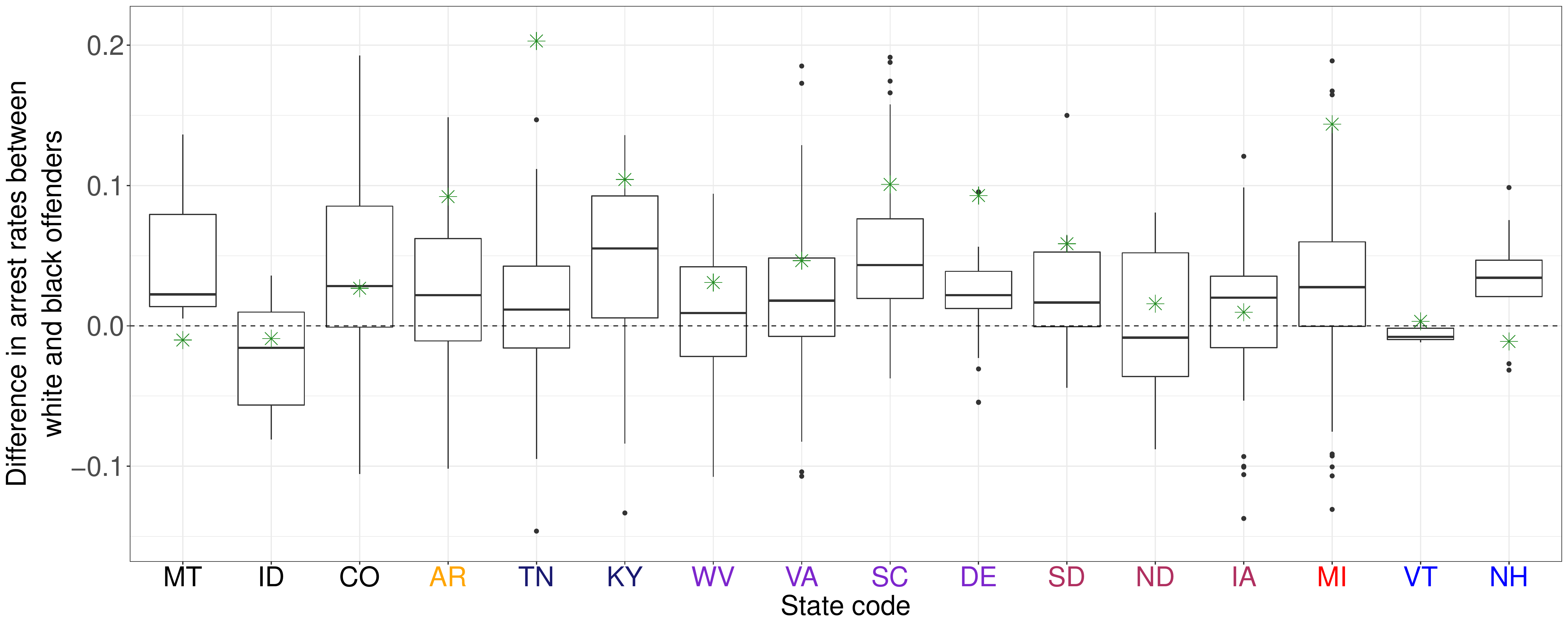}
    \caption{\normalfont{Analysis of racial disparities in arrest rates at the
    level of the states and law enforcement agencies. The green stars indicate
    state-level observed differences in the rates of arrest for simple assaults
    between White and Black offenders in the data from NIBRS considered. For
    example, in Tennessee (TN), the arrest rate is 20\% higher for White offenders than for Black offenders.  States are colored and
    grouped according to the corresponding Census region. The boxplots represent
    the distribution of the gap in arrest rates across police agencies with at
    least 50 Black and White offenders. The lower and upper hinges correspond to
    the first and third quartiles. The upper whiskers extend to the largest
    value no further than 1.5 times the interquantile range from the
    hinge.}}\label{fig:diff_offense_estimate}
\end{figure}

We start by decomposing the gap in arrest rates between White and Black
offenders (13\%=57\%-43\%) into state-level differences, which are represented
by the green stars in Figure \ref{fig:diff_offense_estimate}. Two notable
patterns are worth mentioning. First, there are quite large variations in arrest
rates across states. While in Arkansas less than 35\% of all offenses resulted in
the arrest of the offender, in Delaware and Vermont 79\% and 82\% did respectively.
Second, we observe that the gap in arrest rates between Black and White
offenders varies substantially across states. Although White offenders were arrested at
higher rates than Black offenders in almost all of the states considered, the pooled mean of the gap 
in arrest rates is 6\% (std.dev.=6\%) and thus it is smaller
than the overall gap. 
It is possible, as we shall show, that variations in arrest rates across police agencies within may drive the observed disparities. 
We first focus our analysis on three states with large disparities, including Tennessee (state-level gap=20\%), Michigan (14\%), and South Carolina (10\%), and then turn to the other states.

Let's first consider the case of Tennessee. Our
dataset contains approximately 500,000 offenses of simple assault that occurred
in the state during the period considered. Tennessee contains two large cities,
Memphis and Nashville. In Memphis, 26\% of the offenses resulted in an arrest and
89\% of the offenders were identified as Black. In contrast, in Nashville and
its metropolitan area the arrest rate is 54\% and only 57\% of the offenders
were Black. Although Black and White offenders were arrested at similar rates
within each of the two areas, the aggregate gap in arrest rates between racial
groups is 15\%. Furthermore, data on offenses in Memphis alone accounts for 51\%
of all incidents involving Black offenders in the state. By contrast, only 5\%
of all crimes with White offenders occurred in Memphis. Even if arrest rates largely
vary across the 14 core cities, the gap across racial groups, within each
city, is always small (mean=4\%, std.dev.=3\%). Thus, the low arrest rate
in Memphis, together with the fact that most of the offenses committed by Black
offenders occur in this city, drives the the observed state-level disparity in arrest rates.

Similar findings hold in the case of Michigan. Data from the city of Detroit
contain 17\% of the almost 600,000 simple assault offenses in the state but also
account for 36\% of all incidents involving Black offenders in the state,
compared to only 2\% of those with White offenders. 
White and Black offenders were arrested at similar rates within the city. Since
arrest rates in this area are considerably lower than in the rest of the state
(20\% vs 43\%), the overall disparity in arrest rates at the state level is
large. The case of South Carolina is slightly more complex. For this state we do
not observe large differences in arrest rates between core cities and
surrounding regions and there is no variation in the size of racial disparities
in arrest rates, which are large in both areas. However, we observe that police
agencies in areas where larger shares of offenders are Black also tend to have
lower rates of arrest ($\rho=-0.32$). 

We then study whether the disparities in arrest rates persist at the level of
the police agency for the other states considered. Figure~\ref{fig:diff_offense_estimate} shows the boxplots of
the differences in arrest rates between White and Black offenders for the police agencies in each state. 
We observe that the median gap in arrest rates for the police agencies is generally smaller than the
difference at the state level. This pattern is consistent with
the aforementioned heterogeneity in arrest rates and distributions of the Black
and White populations of offenders across jurisdictions. Interestingly, however, even after
conditioning on the police agency, there still remains some degree of variation
in the gap in arrest rates across racial groups, with arrest rates generally
being higher for White offenders.

By focusing on the characteristics of the law enforcement agency, 
we find that in most states agencies with larger shares
of Black offenders tend to have, as in the case of South Carolina, lower arrest
rates ($\rho=-0.29$), higher crime rates in
their jurisdictions ($\rho=0.37$), more officers per capita ($\rho=0.17$), and
larger populations served ($\rho=0.18$). However, the number of officers per
incident is not linearly associated with the share of offenders that is Black in
the jurisdiction ($\rho=-0.02$, p-value$=0.34$).

We briefly turn our attention to the other four types of offenses. Simple
assault is the least serious crime among the ones that we considered, so ex ante
it is unclear whether the findings would generalize to the other crimes. Based
on our analysis, we can categorize the offenses in two groups. 
The first includes aggravated assault and robbery. For these two types of crimes,
the results are similar to those that we have described above and, more
specifically, arrest rates are higher for White offenders in
almost all of the states considered (Figure~\ref{fig:arrest_rate_by_state}). The
pooled means of the state-level gaps in arrest rates between White and Black
offenders for aggravated assault and robbery are 6\% and 11\% respectively.
Still, we need to keep in mind that the magnitude of arrest rates for these two
types of offenses is very different: While almost two in three offenders were arrested
for aggravated assault (pooled mean=63\%), only one in three reported offenses of robbery resulted
in arrest (35\%). Large differences in arrest rates
across offense types persist even if incidents with multiple
victims and offenders are accounted for. 
For crimes of rape and murder, both the aggregate and state-level differences in arrest rates
across racial groups are small (average state-level diff.=2\% for both types
of offenses). Overall arrest rates are low across all states for rape (pooled mean=29\%) and high for murder (mean=73\%). 

Although these results cannot explain \textit{why} the observed differences in arrest
rates across offenders' racial groups arise and why the appear only for assaults and robberies, they show that the magnitude of disparities can be
explained by the existing large variations in the likelihood of arrest across
jurisdictions and by the fact that agencies with larger shares of Black
offenders have lower arrest rates overall. We discuss two potential
explanations for these findings in \textsection\ref{sec:discussion}.

\subsection{Regression analysis}\label{sec:reg_analysis}

In this section, we investigate whether the observed differences in arrest rates across racial groups 
can be explained by the incident's characteristics. For this purpose, we model
the likelihood of arrest as a function of contextual and situational factors,
fitting separate logistic regression models for each of the five types of violent offenses as described in \textsection\ref{sec:methods}. As a reminder, among the
regressors, we include victim and offender's demographics, the victim-offender relationship prior to the incident, circumstances (e.g., location and time
of the day) and other characteristics (e.g., presence of weapon) of the
incident, information regarding the police agency where the incident was
recorded (police officers per capita, population served, and share of Black
offenders), and the state in which the crime occurred. 
The full set of regressors and coefficients estimates is presented in
Table~\ref{tab:coefficient_regression}.

Consistent with the high-level findings of our macro-level analysis, our estimate of the coefficient of White \textit{offender} race
(vs. Black) is positive and statistically significant in case of aggravated
assault, simple assault, and robbery ($0.03$, $0.04$, $0.24$ respectively), and
not statistically significant for forcible rape and murder ($0.04$ and $-0.2$
respectively). However, note that the coefficients estimates in case of assault,
despite being statistically significant, are close to zero. The coefficient
of White \textit{victim} race is statistically
significant and negative for forcible rape ($-0.11$), and positive for aggravated and simple assault ($0.07$
and $0.09$ respectively). It is not statistically significant for
murder/non-negligent manslaughter and robbery ($-0.02$ and $0.03$
respectively).\footnote{We conducted two additional checks. First, we fitted logistic mixed model with random effects for law enforcement agencies \cite{fitzmaurice2008longitudinal}. 
The estimates of the race coefficients were very close to those of the regression model that we present in the main analysis.
Second, we clustered the sandwich standard errors \cite{freedman2006so} by law enforcement agency and year. The resulting coefficients estimates relative to the offender's race was statistically significant in case of simple assaults, but not for aggravated assaults.}

If the model were correctly specified, we could interpret the coefficient as the
log of the odds ratio of the likelihood of arrest for a White offender
compared to a Black offender, conditional on all other regressors. Our model would then indicate that,
ceteris paribus, an arrest is more likely for a White offender compared to a
Black offender for assaults and robbery, but not for murder and rape. Furthermore, incidents involving White victims would be, ceteris paribus, less likely
to result in an arrest for offenses of rape, but more likely in case
of aggravated and simple assault. The conditional likelihood of arrest does not considerably
differ across racial groups for murder. Note again that, even in the case of several of the statistically significant coefficients, the coefficient estimates themselves are generally small.

More realistically, our logistic regression model is misspecified and thus such
interpretations of the coefficients estimates are not correct. In
Figure~\ref{fig:model_diagnostics}, we present the results of the ``focal
slope'' model diagnostics for the offender's race coefficient across all crimes other than murder/non-negligent manslaughter,
which we omitted due to the small sample size. 
We observe that the estimates of this coefficient vary with the distribution of the regressors, thereby indicating that none of
the regression models is well specified. Interestingly, we observe that the
association of the offender's race (White=1) with the outcome is typically
positive for incidents with White victims, but is weaker or even negative in
case of Black victims. This result would seem to suggest the presence of an
interaction between victim's and offender's race. 
In case of rape and simple assaults with Black victims, 
the offender being White appears to be associated with a decrease in the likelihood of arrest.
Jurisdiction-level factors also appear to impact the
estimates of the offender race coefficient. For example, in case of assaults, we observe
a weak or even negative association between the race variable and the outcome
in police agencies where offenders mainly belong to one racial group, but a
positive association in agencies where the racial composition of offenders is
more diverse. We observe that the association between the offender's race and
arrest is positive and strong only for middle-aged offenders, and is weak for both young and old offenders. Overall, the results indicate that no
causal conclusions can be drawn regarding the ``effect'' of race from such a
model. Furthermore, problematically, the findings around the size of the effect
are contingent upon the areas and types of crimes that are considered.

Finally, although our emphasis was primarily on the coefficient of race, one
should note that most of the coefficient estimates for the other features are
statistically significant and many are much larger than those corresponding to race. This indicates that incidents of the same offense type might result in
different outcomes.

\section{Discussion}\label{sec:discussion}
Our investigation of NIBRS data, which centered on race as a predictor of the
likelihood of arrest for violent crimes, was motivated by the implications
concerning the potential disconnect between re-offense and re-arrest in RAIs. 
We presented two key findings.
First, in our analysis White offenders were more likely to be
arrested than Black offenders for crimes of robbery and assaults, but not for
forcible rape and murder/non-negligent manslaughter.  The finding that arrest rates in the NIBRS are higher for White offenders than
for Black offenders was also noted by~\citet{d2003race}. The observed disparities in our study
are largely explained by variations in arrest rates across jurisdictions. 
Yet, even after conditioning on the individual law enforcement agency, some of 
these disparities persisted. 
Second, while our initial regression results indicated that the magnitude of the offender race coefficients estimates were fairly small compared to those of other
predictors, the model diagnostics revealed that their sign and magnitude varied with the sample of offenses that was considered.
Despite the misspecification of our regression model, it seems unlikely that crime characteristics that we took into account could fully explain the observed variations or even the racial disparities in arrest rates. 
Our results call into question the reliability of arrest as a proxy for offense. 

\paragraph{Arrest rates and racial disparities vary considerably across jurisdictions} 
In our study we also found that the observed racial disparity in arrest rates can be attributed to lower arrest rates in police agencies with larger shares of Black offenders.  This finding is aligned with several hypotheses from the criminology literature, two of which we discuss below.  

The first explanation builds off of the hypothesis of benign neglect, which posits that the effort
invested by police into the investigation of the crimes might be proportional to the
perception of the victim's deservedness. A perspective of victim devaluing,
which is more likely to prevail in disadvantaged neighborhoods or cities, might,
in combination with resource constraints, lead to systematically lower 
investigation effort devoted to incidents occurring in those areas \citep{stark1987deviant}.
In particular, police departments might allocate their limited resources only to
certain communities and to solving the most serious incidents
\citep{klinger1997negotiating}, such as rape and murder, for which police
officers are granted less discretion. Police might also be more likely
to tolerate criminal acts in areas where crimes rates are higher
\citep{stark1987deviant,klinger1997negotiating} or that are racially
segregated~\cite{kent2005minority}. It would then be hard for victims and
residents in these areas to be heard by the police and to legitimate their
complaints. Consistent with Black's stratification hypothesis
\citep{Black1976behavior}, which posits a positive relationship between the
likelihood of arrest and the gap in social status of victim and offender, this
explanation is partially supported by our finding that the association between
the probability of arrest and the offender White race is weaker or negative for incidents involving Black
victims. By contrast, the investigation of crimes of murder and rape would
be less affected by the differential allocation of policing resources. 
It is possible that the results for murder incidents 
may reflect a shift in racial discrimination from a victim- to an offender-centered
perspective: Murders are the most serious form of crime and as such they attract
considerable attention from the media, community, and police agencies. In
addition, the evaluation of performance based exclusively on the clearance rate
represents an incentive for homicide detectives to solve all cases to which  they are
assigned \cite{puckett2003factors}. 

A second hypothesis is that the observed racial
disparities are an artefact of variations in the community's willingness to rely
and cooperate with the police. In the case of violent crimes, the cooperation of
victims and onlookers with law enforcement is often necessary for identifying
the suspect and clearing the crime. Certain communities have lower levels of trust in law enforcement---a phenomenon known as legal cynicism---and consequently
are less likely to assist police officers in their investigations. For example,
social norms can amplify negative attitudes towards the police by incentivizing
the resolution of conflicts without seeking help from members outside the
community \citep{anderson2000code} or by fostering a ``stop snitching'' culture
\citep{clampet2015sliding}. Community members may also fear retaliation if they
decide cooperate with the police \citep{kubrin2003retaliatory}. There is considerable evidence that cooperation varies with these types of (sub)cultural factors
\citep{sunshine2003role,kirk2011legal,carr2007we,tyler2008legitimacy,tyler2014street}.
Phenomena that erode trust, such as police misconduct
\citep{terrill2003neighborhood,kane2002social} and over-policing
\citep{kane2005compromised} (e.g., street stops) are more frequent in
predominantly-Black areas. Consistently, many studies have found Black
individuals to hold more negative views of law enforcement
\citep{weitzer1999race,weitzer2004race,weitzer2015policing}. 
Differential cooperation has been a central theme in
prior work on racial disparities in arrests for homicides
\citep{puckett2003factors, fagan2018police}. 
Unfortunately, one clear limitation of NIBRS data is that they do not include incident-level
details on cooperation, as also \citet{roberts2016crime} noted in their recent study.  
The only piece of information that is available, and that has been analysed by
recent work on violent offenses, is whether the incident is exceptionally
cleared because of the lack of victim's cooperation \citep{felson2016victims}.
In our dataset, however, the share of incidents that were cleared for this
reason did not differ across racial groups of offenders and victims (see
Appendix \ref{sec:additional_data}).

The two hypotheses should not be seen as a dichotomy: Each of them may explain, at least in part, the observed disparities in the probability of
arrest. At the same time, it also seems certainly possible that these
disparities could be explained by differences in the ways in
which crimes are recorded across law enforcement agencies or by different legal
standards used to evaluate whether an arrest needs to be made. For example, in the 
case of intimate partner violence (a.k.a. IPV), once the incident is reported to the police, prevailing mandatory arrest laws in certain jurisdictions leave officers with little discretion as to whether to arrest the reported offender
 \citep{hirschel2007domestic,dugan2003domestic,kochel2011effect}. This phenomenon might partially explain the
variation observed in arrest rates for simple assaults, which include some IPV offenses, across states. 

\paragraph{Reporting a single estimate of the coefficient of the offender's race may be misleading} 
Our focal slope model diagnostics of the logistic regression revealed that the
association between the offender's race and the likelihood of arrest largely
depended on the sample of offenses that was considered. This result is
important for two reasons. First, it demonstrates that the regression model
employed in our analysis is not well-specified. In past studies, researchers
typically postulated a model, fitted it on the data, and then reported the
estimates of the coefficients of interests, such as those of the offender's or
victim's races. 
The models used in these works were often similar to ours. For
example, \citet{d2003race} employed a logistic regression that accounted for a
subset of the predictors that were included in our model. The focal slope
diagnostics in our analysis, which are based on the recent work of
\citet{buja2019models}, have shown that reporting one single coefficients
estimate may not be very informative when the modeling assumptions do not hold.
Researchers should, instead, analyse how the misspecification of the regression
model that they chose may call into question their interpretations of the
results, e.g., by using the types of diagnostics proposed by \citet{buja2019models}.
Second, the diagnostics have revealed that, after controlling for crimes characteristics, the association between the likelihood of arrest and the offender's race depended on the race of the victim. 
We also observed a weaker association between
the offender's race and the likelihood of arrest in police agencies with larger
shares of White offenders. While a careful analysis of the role of race as a
predictor of the likelihood of arrest across each subpopulation (e.g., different
jurisdictions) is challenging and outside of the scope of our work, the main
takeaway from our results is that the likelihood of arrest can differ across
races, both at the aggregate level and conditionally on other factors. 

\paragraph{Arrest is a biased proxy of offense} 
In \textsection\ref{sec:intro}, we discussed how prior studies have found that the racial composition of arrestees in
the Uniform Crime Reporting (UCR) roughly match that of offenders in the National Crime
Victimization Survey (NCVS) data, a criminal victimization survey based on a
nationally representative sample of households, for the four types of
violent crime measured by both programs. 
This pattern suggests that the likelihood of arrest for White and Black
offenders is, on average, similar, but it does not indicate that the likelihood of arrest is equal for {\it all} offenders in the two racial groups.
We have found that arrest rates and racial disparities vary across states and jurisdictions.
The regression analysis has further revealed that, even after conditioning on relevant features, the disparities across racial groups still persist and run in different directions depending on these features. 
Thus, the extent to which arrest reflects offending behavior depends on the geographical area where the crime occurs, the offenders' and victims' demographics, and other crimes characteristics.
Treating arrests as an unbiased, or even racially-unbiased, measure of violent offending
fails to acknowledge the existing heterogeneity in the likelihood of
arrest.
\section{Implications for RAIs} 
\paragraph{Risk of re-arrest and re-offense may diverge, even in case of violent offenses.  } Our analysis has shown that arrests are not a random sample of all
violent offenses known to police agencies as reflected in NIBRS data. The large
variations that we observe in arrest rates across crime types and jurisdictions
are deeply problematic for RAIs.\footnote{We acknowledge that availability of
information regarding the offender and victim might influence the likelihood of
arrest differently across crime types. However, the observed variation persisted even once 
we considered all incidents in the data, i.e., we only applied the data restrictions relative the states. The arrest rates for 
murder, forcible rape, robbery, aggravated assault, and simple assault were 
58\%, 24\%, 24\%, 50\%, and 50\% respectively.} As an example, let's revisit our analysis of Tennessee's offense
and arrest data. In Nashville, half of all simple assault offenses resulted in
arrests, as compared to only one fourth in Memphis. We should expect these
differences to be reflected in the prevalence of re-arrests in the data used to
train and assess RAIs. The variation in the characteristics of the offender
populations would then shape the predictors used to construct the RAI. Consequently,
we should expect RAIs trained on re-arrest data to underestimate the risk of re-offense for
certain subpopulations of defendants---specifically, those that are more prevalent in the city with lower arrest rates, and to be poorly calibrated with respect to re-offense data. Moreover, if the risk factors employed
by the RAI cannot explain the disparities observed between the two cities, the
instrument will also exhibit differences in its calibration properties on
re-arrest data across the two cities. 
Absent knowledge of the functional dependency
between re-arrest and re-offense, practitioners might try to mitigate the
predictive biases of the RAI using only re-arrest data, only inadvertently
(potentially) exacerbating the bias as measured with respect to re-offense. To overcome this issue,
individual jurisdictions could construct their own RAIs. However, in most cases
this is certainly not possible due to data unavailability and other limitations.

As a second example, consider the variation in arrest rates across types of
offenses. If the propensity to commit different types of violent crime were
to differ across offenders, then the risk of re-arrest would be lower for
certain individuals, even if their risk of re-offense were identical. For
instance, since arrests for rape are rare (as is generally true for sex
offenses), RAIs might erroneously underestimate the risk of these sex offenders
recidivating (compared to, say, aggravated assault offenders), and could nudge
nudge decision-makers toward more lenient treatment that is unwarranted on the basis of underlying re-offense risk.  
In order to identify defendants at high risk of recidivism, practitioners should
design RAIs that either account for this bias or that target only the particular type of crime of interest.
We note that there already exist tools that predict specifically the likelihood of future IPV \cite{messing2013average}.

The consequences of the sampling bias in arrest data are not limited to
mismeasurement of re-arrest outcomes on which RAIs are trained. This bias
affects offenders' criminal histories as well, making them inaccurate
reflections of past offending behavior. The number of prior arrests represents
an underestimation of the total number of prior offenses, especially for
frequent offenders, who are generally less likely to be arrested
\citep{blumstein2010linking}. Similarly, our analysis has shown that certain
types of crimes are more likely to result in arrest, and thus the
characteristics of the crimes recorded in a defendant's prior criminal history
do not offer an unbiased picture of the characteristics of the prior offenses
that they actually have committed. Lastly, the sample of arrestees does not
reflect the sample of offenders in the population. In particular, more frequent
offenders may be less likely to be caught for each individual crime that they
commit, yet more likely to be represented among arrestees, e.g., see the
phenomenon of ``stochastic selectivity'' described by \citet{canela1997relationship}.

\paragraph{Bias cannot be estimated from NIBRS data alone} 
NIBRS is the main program of crime data collection in the US, providing the best
available unified view of crime trends. Large-scale analyses of racial
disparities could leverage the NIBRS because of its comprehensiveness. However,
for the purpose of mitigating bias in criminal RAIs, NIBRS data alone are
not sufficient for at least two key reasons, which we discuss below.

First, the effect of many types of sampling biases cannot be identified from the observed data. In the
discussion, we have hypothesized that variations in the likelihood of arrest
might stem from multiple sources, namely the allocation of policing resources,
community's cooperation, and ways in which the crimes are recorded. However, it is
also clear that some of the differences in arrest rates across crimes are
inherently tied to the difficulty of identifying and arresting the offender. For
example, crime incidents in which the offender is a stranger to the victim are
cleared at lower rates across all crime types
(Table~\ref{tab:coefficient_regression}). Many of these factors are not
represented in NIBRS data, which offers only a limited window into the reason why a reported offense did not result in arrest (e.g., for exceptional clearances). 
In addition,
NIBRS data only include crime incidents that are known to law enforcement, which
represent only a small share of the offenses that are committed. According
to estimates from the National Crime Victimization Survey (NCVS), in 2019 only about
two in five violent victimizations were reported to the police
\citep{morgan2019criminal}. Evidence suggests that rates of crime reporting
substantially vary across crime types \citep{baumer2010reporting} and are
generally higher in case of incidents involving Black victims, even if cooperation rates may be
lower \citep{xie2012racial}. Thus, the information present in NIBRS data would
not be sufficient to implement tailored ``debiasing'' approaches that target
specific types of bias in the dat (e.g., a certain discriminatory proactive
policing strategy) and that may be necessary from a legal
perspective~\cite{ho2020affirmative}. One could, however, attempt to leverage
NIBRS together with richer datasets of police records, of self-reported
offending behavior, data from NCVS, and of citizen-police interactions such as
those studied in~\citet{terrill2003neighborhood}. \citet{kochel2011effect} and
\citet{lytle2014effects} offer interesting examples of meta-analyses on the
role of race in arrests for different types of crime types and data.
An additional potential limitation is the issue of model misspecification that
we encountered in our analysis. Clearly, regression models more complex than the
ones that we employed that could potentially address this issue, but they
may be harder to interpret and sanity-check as corrective mechanisms. It is important to keep such trade-offs in mind when considering model-based bias mitigation strategies. 

Second, static estimates of the sampling bias may not be valid for longitudinal
outcomes. Due to the longitudinal nature of the measurement of re-arrest
outcomes on which RAIs are trained and assessed, knowledge of the offending
frequency and of the characteristics of the crimes committed would be needed for
the correction. Data from NIBRS do not contain offenders' identifiers and thus
cannot provide such estimates. However, offending behavior has been studied by
criminologists by surveying inmates and through longitudinal studies for decades. M
ost studies agree that the offending frequency varies
across populations and, while most of the offenders commit only a few crimes a
year \citep{piquero2007does}, there is a small group of offenders with very high
offending rates (more than 100 crimes a year) for whom the  probability of
arrest is low~\cite{blumstein2010linking}. Attempts to correct for bias could
then draw on these findings. 

We remark that the utility of data from NIBRS for the estimation of sampling
bias in RAIs seems fairly limited.  Alternative sources of data, as discussed above, may provide more promising avenues to bias mitigation.  However, what is gained in specificity with alternative data sources often comes at the cost of generalizability. For example,
data on self-reported offending behavior has mainly been collected for
populations of youths living in certain geographical regions, such as the
Pathways to Desistance longitudinal study conducted on young adolescents in
Philadelphia (Pennsylvania) and Maricopa county
(Arizona) \cite{brame2004criminal}. Extrapolation of findings on these samples
to other populations, such as offenders that are older or in other geographical
areas, would need to be carefully considered.

\section{Conclusions}

Characterizing the biases present in arrest data 
is critical to assessing the challenges and opportunities posed by recidivism risk assessment instruments. 
While past work has suggested evaluating RAIs on data of violent offenses 
as a way to alleviate concerns of racial bias, 
by analyzing incident-level crime data from NIBRS, 
we have shown that violent offense data 
suffer from significant sampling biases. 
There are disparities in the probability of arrest across geographical areas,
characteristics of the crimes, 
and racial groups of victims and offenders, 
that were not adequately captured in prior work.
Our analysis indicates that arrest data on violent crimes 
are not a random sample of the offenses actually committed. 
Absent corrective mechanisms, the predictions produced by RAIs trained on such data can be expected to reflect these sampling biases.  
At the same time, 
correcting for the sampling bias
in recorded arrest data 
without accounting for the complex relationship between race, offending, and arrest
may exacerbate the disparities
exhibited by the RAI. 
In the present work we have sought to highlight several of the key complexities and challenges 
of identifying the magnitude of sampling bias in arrest data.  We also discussed the limitations and challenges of using various sources of offending and arrest data in bias mitigation efforts.  Despite these challenges, we believe it is important for future work on developing and de-biasing recidivism risk assessment instruments to explicitly consider the discrepancy between criminal offending and arrest.  


\section{Acknowledgments}
We are grateful to the Partnership on AI (PAI) and the Carnegie Mellon University Digital Transformation and Innovation Center sponsored by PwC for funding this research.
We also thank Arun Kumar Kuchibhotla and anonymous reviewers for providing valuable feedback. 

\bibliographystyle{ACM-Reference-Format}
\bibliography{references}

\appendix

\section{Additional details on data analysis}\label{sec:additional_data}

\paragraph{Clearance by arrest and by exceptional means}

Incidents can be cleared either by an arrest or by exceptional means. An offense
is cleared by arrest when at least one person is arrested, charged with the
commission of the offense, and turned over to the court for
prosecution~\cite{federal2004uniform}. In case of individuals below the age of
18, physical arrest is not necessary. The arrest of multiple offenders may
correspond to the clearance of one offense, and many offenses may be cleared by
only one arrest. An offense is cleared by exceptional means if the following
conditions are satisfied: (i) the investigation has established the identity of
the offender; (ii) there is enough information to support an arrest, charge, and
turn over the offender to the court for prosecution; (iii) the exact location of
the offender is known and the offender could be taken into custody; (iv) there
is some reason outside of law enforcement control that precludes arresting,
charging, and prosecuting the offender. Clearance by exceptional means occurs in
case of death of the offender, denied prosecution, custody of other
jurisdiction, refusal of the victim to cooperate, when the offender is a
juvenile and thus they are not taken into custody.

We investigate whether there exist differences in the likelihood of clearance by
exceptiona means and reasons behind the clearance across racial groups of
offenders and victims. The type of offense with the largest share of incidents
that are cleared by exceptional means is forcible rape (about 20\%). Slightly
more than 10\% of these incidents are cleared because of prosecutorial
declination to prosecute and another 8\% due to lack of
victim's cooperation. In case of murder/non-negligent manslaughter, 5\% and 14\% of all
incidents are cleared by exceptional means for Black and White offenders. The
majority of these offenses results into the death of the offender. Clearance by
exceptional means in offenses of robbery (8\%), aggravated assaults (9\%), and simple assaults (14\%) is mainly due to
prosecutorial declination to prosecute and victims' lack of cooperation, the
latter likely being due to cases of domestic violence. Incidents are cleared at
the same rates across the two racial groups for all crime types other than
murder/non-negligent manslaughter. 
An analogous analysis of the likelihood of exceptional clearance 
conditioning on the racial groups of the victims reveals similar results.



\begin{figure*}[b]
  \centering
  \begin{subfigure}{.45\textwidth}
  \centering
  \includegraphics[width=\linewidth]{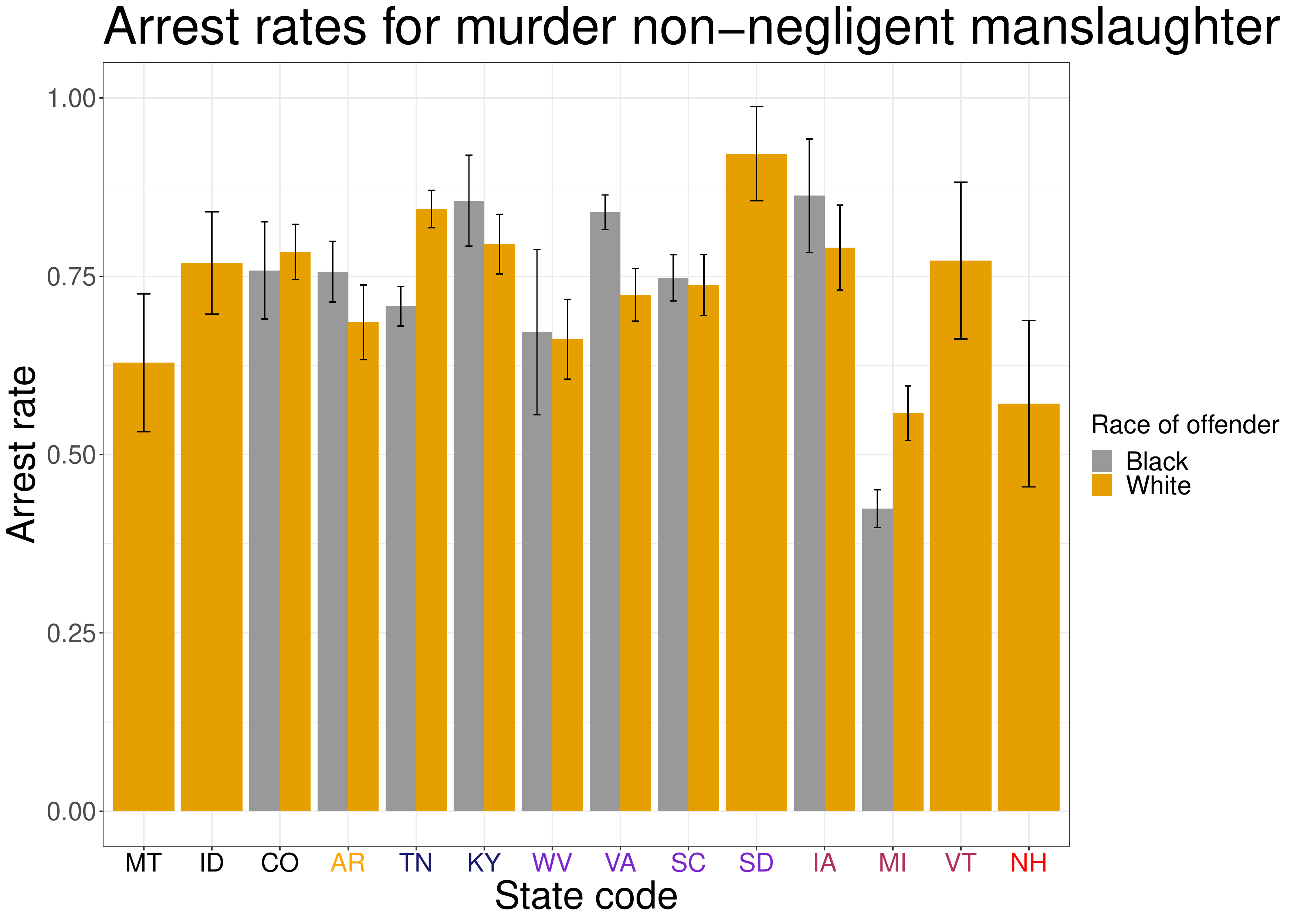}
  \end{subfigure}%
  \begin{subfigure}{.45\textwidth}
  \centering
  \includegraphics[width=\linewidth]{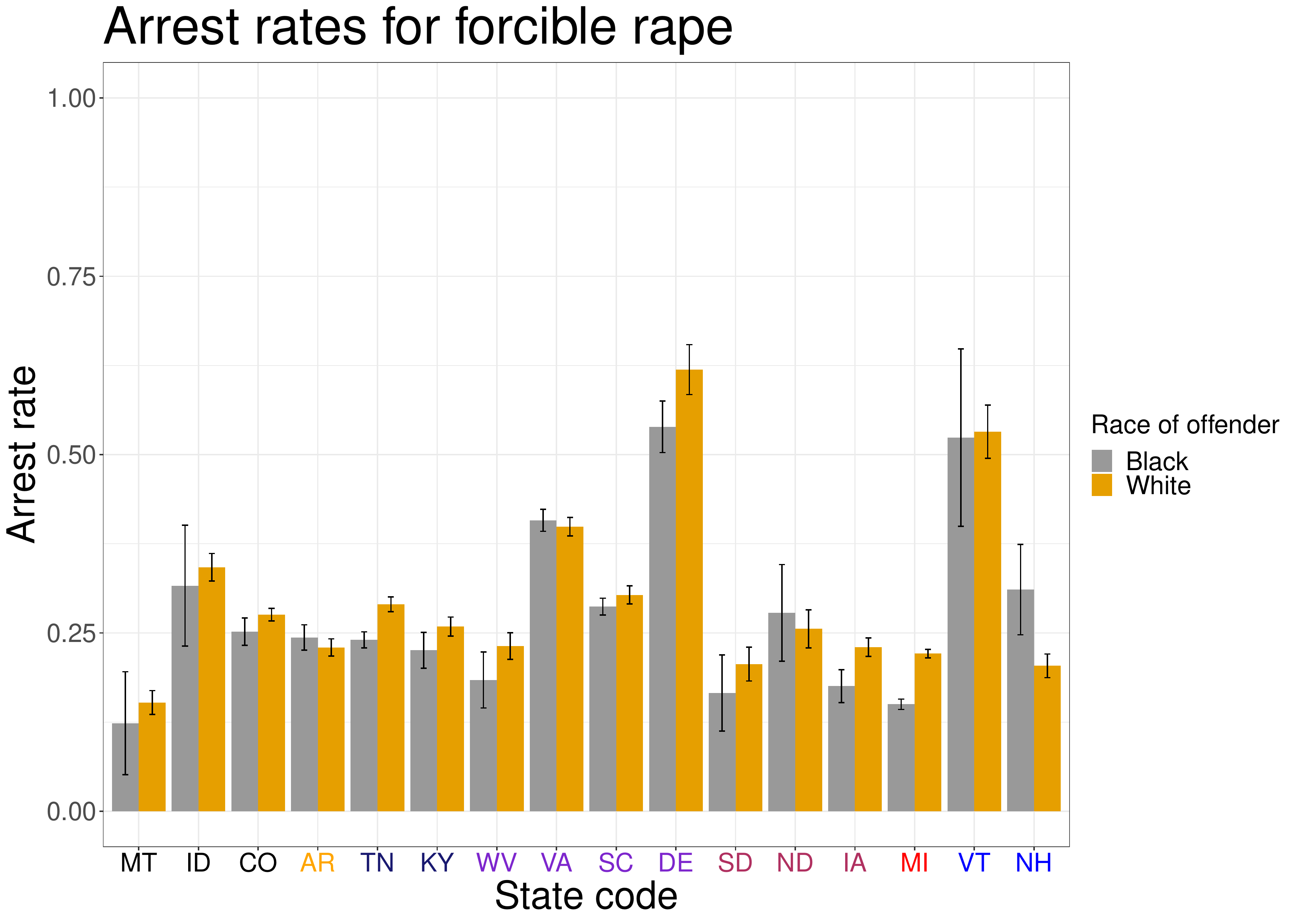}
  \end{subfigure}%
  
  \begin{subfigure}{.45\textwidth}
  \centering
  \includegraphics[width=\linewidth]{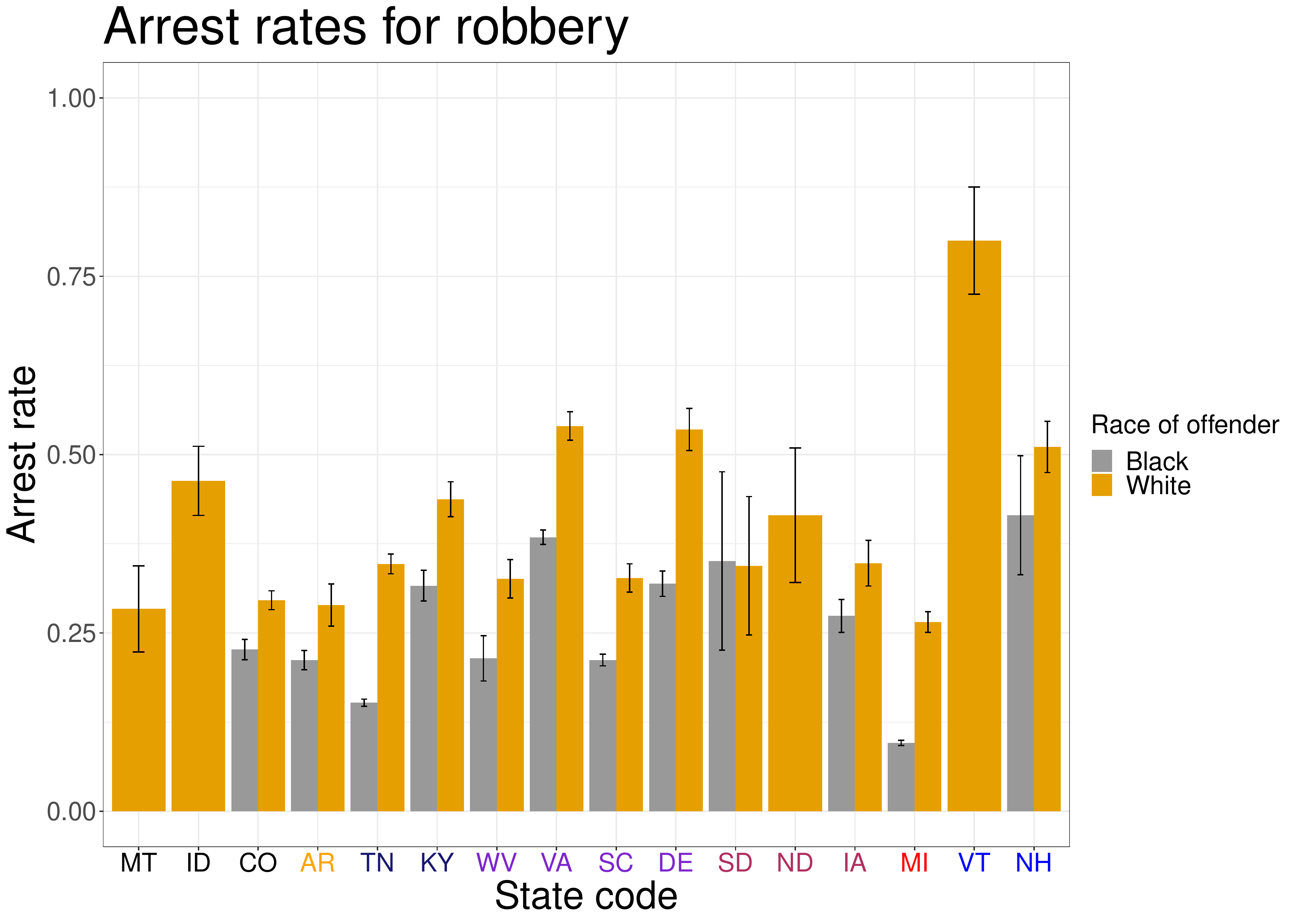}
  \end{subfigure}%
  \begin{subfigure}{.45\textwidth}
  \centering
  \includegraphics[width=\linewidth]{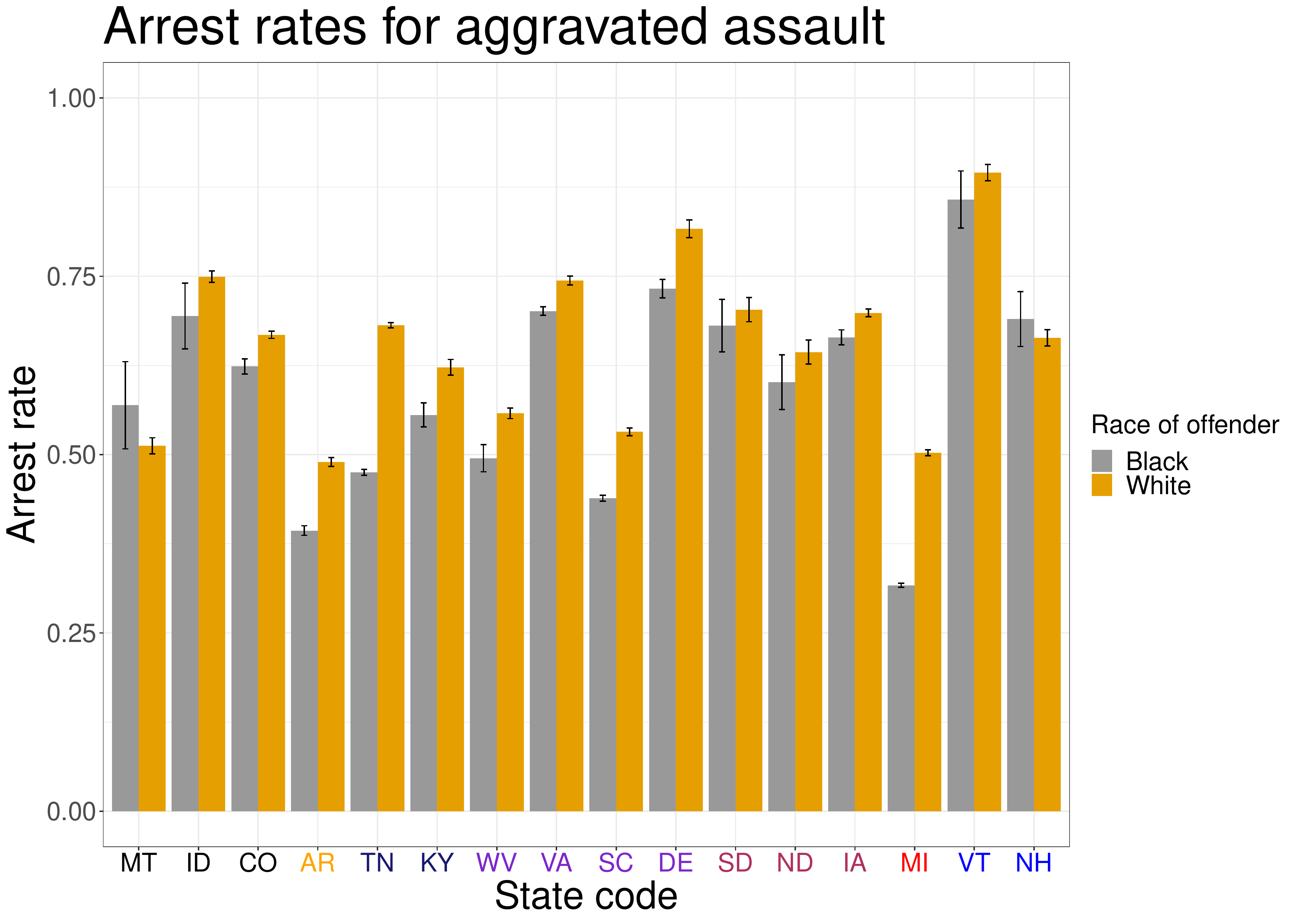}
  \end{subfigure}%
  \caption{\normalfont{Arrest rates by offender's race and state for murder/non-negligent
  manslaughter, forcible rape, robbery, and aggravated assault in the NIBRS data
  considered in the analysis. Only groups of observations (state-offender's race) containing more than 50 crime incidents
  are considered. Error bars indicate 95\% confidence intervals. Note that arrest rates largely vary across crime types and states. They are highest in case of murder/non-negligent manslaughter, and lowest for
  robbery and forcible rape. The gap in arrest rates between offenders' racial
  groups within states is generally small.}}\label{fig:arrest_rate_by_state}
\end{figure*}

\begin{figure*}
  \centering
  \includegraphics[width=0.9\linewidth,  keepaspectratio]{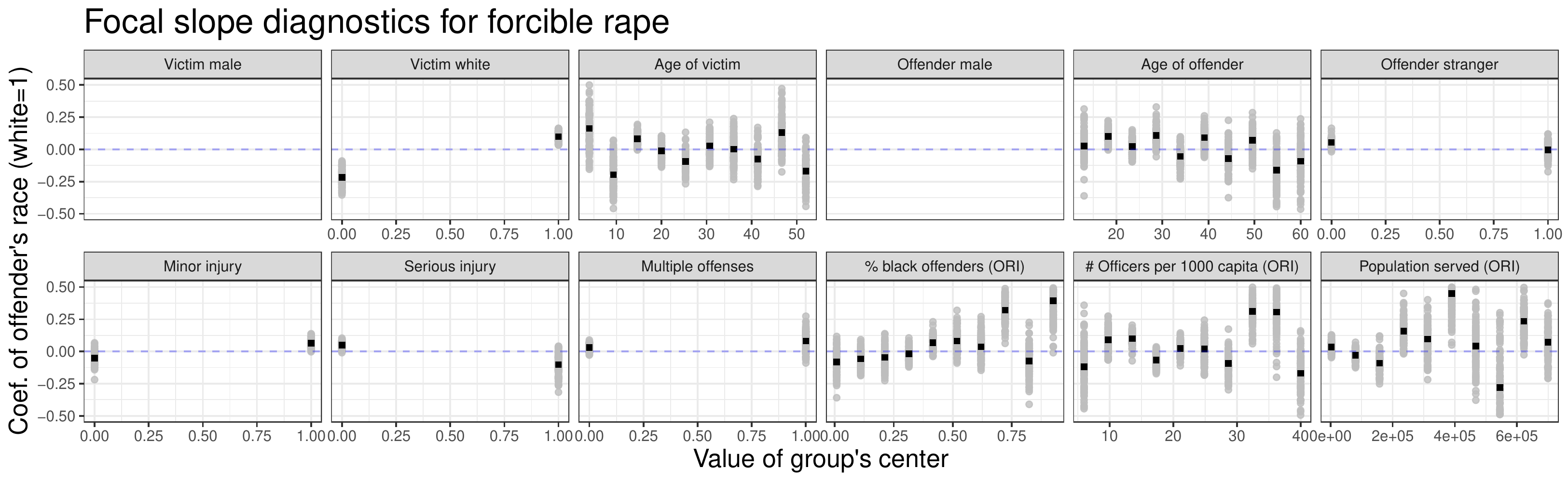}
  \includegraphics[width=0.9\linewidth, keepaspectratio]{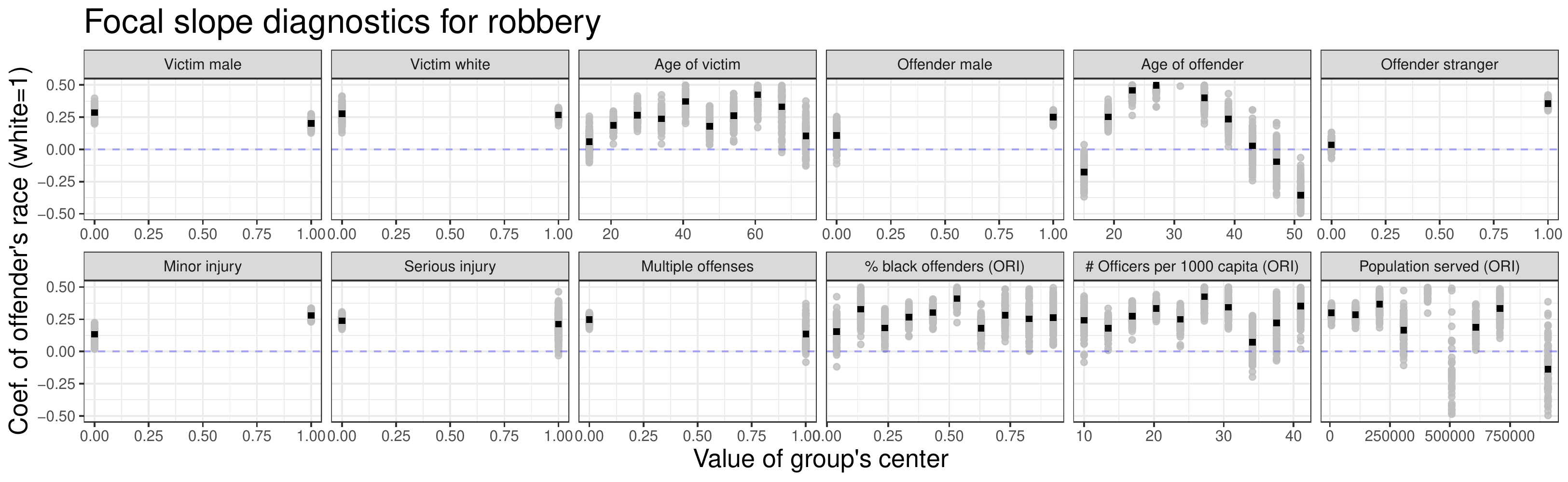}
  \includegraphics[width=0.9\linewidth, keepaspectratio]{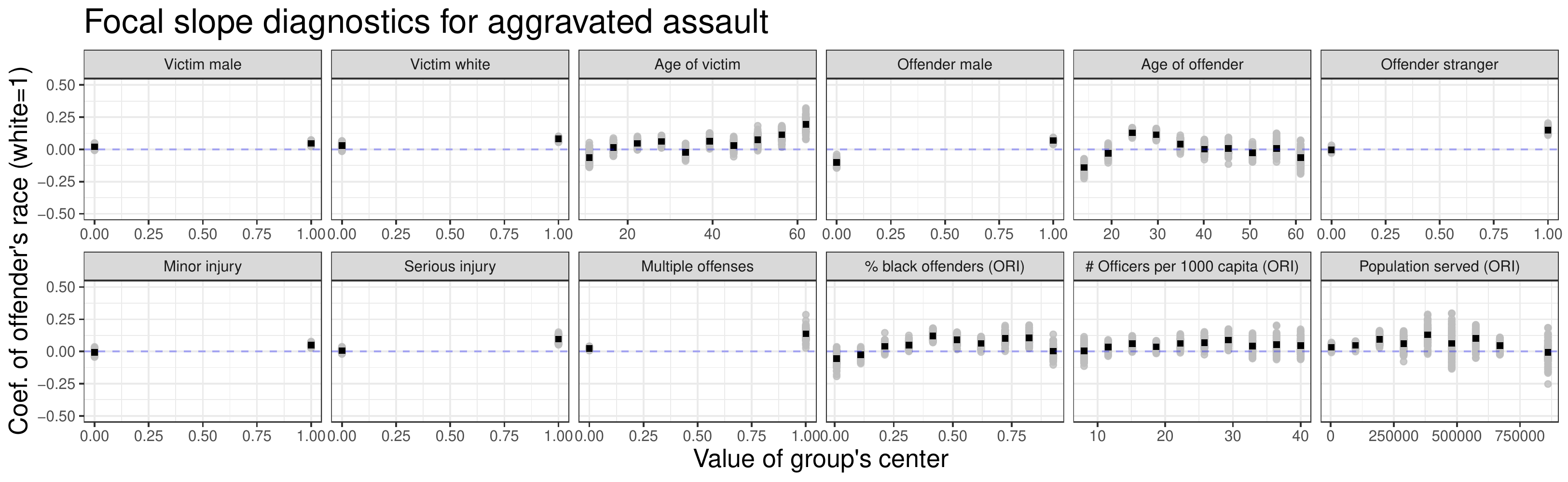}
  \includegraphics[width=0.9\linewidth,  keepaspectratio]{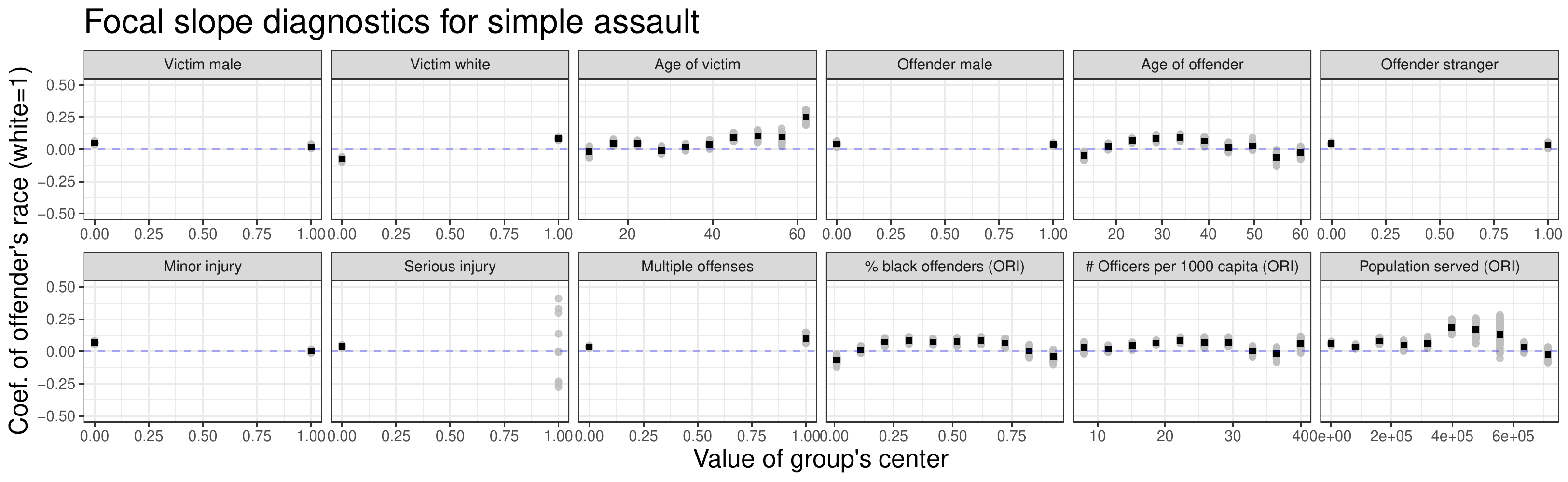}
  \caption{{\normalfont ``Focal slope'' model diagnostic for the logistic regression model whose coefficients estimates are presented in
  table~\ref{tab:coefficient_regression}. The diagnostics for the offense of
  murder/non-negligent manslaughter are omitted due to the small sample size. 
  The methodology behind the diagnostics is described in \textsection\ref{sec:methods}.
  The grey points correspond to the coefficient estimates relative to the offender's
  race (White=1) obtained by fitting the logistic regression model on each of 100
  bootstrapped datasets, for each feature (panel's title) and feature's grid value
  (value on the grid, horizontal axis). The black dots correspond to the means of
  such estimates. 
  We observe that the size and sign of the values of the black dots vary across the range of
  the regressors. This suggests the presence of interactions between race and the
  regressors, thereby indicating the misspecification of our modeling approach.  
  For example, the association between the offender's race (White=1) and the
  outcome is positive when the victim is White, but it's close to zero or even negative
  when the victim's race is Black across three of the four crime types. In case of assaults,
  the association is weak or even negative in police agencies where offenders
  mainly belong to one of the two racial groups considered, and positive
  otherwise. 
  }}\label{fig:model_diagnostics}
\end{figure*}

\paragraph{Additional results for \textsection\ref{sec:results}.}
Figure~\ref{fig:arrest_rate_by_state} shows the arrest rates by state and
offender's race. 
The coefficients estimates of the logistic regression model are presented in Table \ref{tab:coefficient_regression}.
Lastly, Figure \ref{fig:model_diagnostics} displays the focal slope model diagnostics. 
More details regarding these
figures are contained in the captions and in the main body of the paper.

\paragraph{Software} In our analysis, we used the following \texttt{R} packages.
For data processing and visualization, we used the \texttt{tidyverse} set of
packages \cite{tidyverse}. Specifically for the statistical modeling, we used
\texttt{fastglm} \cite{fastglm}, \texttt{lme4} \cite{lme4}, \texttt{mice}
\cite{micepkg}, 
\texttt{sandwich} \cite{zeileis2020various}. In addition, we used \texttt{icpsrdata} \cite{icpsrdata},
\texttt{asciiSetupReader} \cite{asciiSetupReader}, \texttt{vroom} \cite{vroomR},
\texttt{haven} \cite{haven}, \texttt{renv} \cite{renv}, \texttt{knitr}
\cite{xie2014knitr}, \texttt{kableExtra} \cite{kableextra}, \texttt{xtable}
\cite{xtable}, \texttt{cli} \cite{cliR}, \texttt{here} \cite{hereR},
\texttt{furrr} \cite{furrr}.

\begin{table*}[t] 
  \caption{{\normalfont Summary statistics of NIBRS data of violent offenses considered in the analysis,
   grouped by the offender's race}}\label{tab:summary_stats_data} \centering
   \resizebox{0.8\textwidth}{!}{ 
   \begin{tabular}{lcccccccccc} \textbf{Variable}
   & \multicolumn{2}{c}{\textbf{Murder/n.n.m.}} &
   \multicolumn{2}{c}{\textbf{Forcible rape}} &
   \multicolumn{2}{c}{\textbf{Robbery}} &
   \multicolumn{2}{c}{\textbf{Aggravated assault}} &
   \multicolumn{2}{c}{\textbf{Simple assault}} \\
   \toprule Race of offender & Black & White & Black & White & Black & White &
   Black & White & Black & White \\
   \midrule \# observations & 4787 & 4394 & 32532 & 70777 & 76566 & 24567 &
   275697 & 320627 & 1045225 & 1624174 \\ \% observations & 52\% & 48\% & 31\%
   & 69\% & 76\% & 24\% & 46\% & 54\% & 39\% & 61\% \\  \midrule \%
   intraracial crime & 86\% & 94\% & 63\% & 97\% & 57\% & 91\% & 81\% & 94\% &
   78\% & 96\% \\
     \midrule \% arrests & 67\% & 73\% & 25\% & 27\% & 19\% & 36\% & 44\% &
   62\% & 43\% & 57\% \\ \% arrests (intraracial) & 66\% & 73\% & 27\% & 27\%
   & 17\% & 37\% & 43\% & 62\% & 42\% & 58\% \\
   \% arrests (interracial) & 71\% & 71\% & 22\% & 23\% & 21\% & 30\% & 51\% &
   51\% & 49\% & 47\% \\ \bottomrule \end{tabular}
   }

   {\raggedright {\footnotesize Notes: ``\# observations'' and ``\% observations'' indicate the
   number of observation 
   and the ratio between the number of offender of a
   certain race and the total number of offenders for that crime respectively in each bucket (e.g., Black simple assault offenders). ``\%
   intraracial crime'' corresponds to the share of victims whose race is the
   same as the offender's race, i.e., intraracial crime. ``\% arrests'', ``\%
   arrests (intraracial)'', ``\% arrests (interracial)'' indicate the arrest
   rates for all offenses, only intraracial offenses, and only interracial
   offenses (conditioning on the offender's race) respectively.} \par}
\end{table*}

\begin{table*}[t]
  \caption{{\normalfont Coefficients estimates (and sandwich standard errors) of
  a logistic regression model that estimates the likelihood of arrest on NIBRS
  data}}\label{tab:coefficient_regression}
  \centering
  \resizebox{0.85\textwidth}{!}{\begin{tabular}{llllll}
   \textbf{Variable} & \textbf{Murder/n.n.m.} & \textbf{Forcible rape} &
   \textbf{Robbery} & \textbf{Aggravated assault} & \textbf{Simple{} assault} \\ 
    \toprule
  Age of offender & -0.01 (0.00)*** & \phantom{-}0.01 (0.00)*** & \phantom{-}0.01 (0.00)*** & \phantom{-}0.00 (0.00)*** & -0.01 (0.00)*** \\ 
  Offender male & -0.18 (0.09)*   & \phantom{-}2.74 (1.02)**  & -0.11 (0.03)**  & -0.11 (0.01)*** & \phantom{-}0.01 (0.00)*** \\ 
  Offender white & -0.20 (0.09)*   & \phantom{-}0.04 (0.02).   & \phantom{-}0.24 (0.02)*** & \phantom{-}0.03 (0.01)*** & \phantom{-}0.04 (0.00)*** \\ 
  Age of victim & \phantom{-}0.00 (0.00)    & -0.02 (0.00)*** & \phantom{-}0.00 (0.00)    & \phantom{-}0.01 (0.00)*** & \phantom{-}0.01 (0.00)*** \\ 
  Victim male & \phantom{-}0.20 (0.06)*** & \phantom{-}2.64 (1.02)**  & -0.10 (0.02)*** & -0.12 (0.01)*** & -0.14 (0.00)*** \\ 
  Victim white & -0.02 (0.08)    & -0.11 (0.03)*** & \phantom{-}0.03 (0.02)    & \phantom{-}0.07 (0.01)*** & \phantom{-}0.09 (0.00)*** \\ 
  Minor injury &  & -0.31 (0.02)*** & \phantom{-}0.01 (0.02)    & -0.24 (0.01)*** & -0.50 (0.00)*** \\ 
  Serious injury &  & \phantom{-}0.32 (0.03)*** & \phantom{-}0.16 (0.03)*** & \phantom{-}0.25 (0.01)*** & \phantom{-}0.83 (0.38)*   \\ 
  During day & -0.01 (0.05)    & -0.10 (0.02)*** & \phantom{-}0.31 (0.02)*** & -0.02 (0.01)*** & -0.11 (0.00)*** \\ 
  Offender stranger & -0.48 (0.06)*** & -0.66 (0.02)*** & -0.81 (0.02)*** & -0.53 (0.01)*** & -0.32 (0.00)*** \\ 
  Residence & \phantom{-}0.18 (0.05)*** & \phantom{-}0.20 (0.02)*** & -0.08 (0.02)*** & \phantom{-}0.32 (0.01)*** & \phantom{-}0.30 (0.00)*** \\ 
  Offender alcohol use & \phantom{-}0.63 (0.11)*** & \phantom{-}0.01 (0.02)    & \phantom{-}0.69 (0.04)*** & \phantom{-}0.56 (0.01)*** & \phantom{-}0.56 (0.00)*** \\ 
  Offender substance use & \phantom{-}0.02 (0.14)    & -0.36 (0.04)*** & \phantom{-}0.22 (0.06)*** & -0.02 (0.02)    & \phantom{-}0.09 (0.01)*** \\ 
  Firearm present & -0.58 (0.08)*** & -0.16 (0.07)*   & -0.27 (0.02)*** & -0.24 (0.01)*** &  \\ 
  Other weapon present & \phantom{-}0.14 (0.09)    & \phantom{-}0.15 (0.05)**  & -0.02 (0.03)    & \phantom{-}0.09 (0.01)*** &  \\ 
  Offense not completed &  & \phantom{-}0.16 (0.03)*** & \phantom{-}0.11 (0.02)*** &  &  \\ 
  \# Officers per 1000 capita (ORI) & \phantom{-}0.00 (0.00)    & \phantom{-}0.01 (0.00)*** & \phantom{-}0.00 (0.00)    & \phantom{-}0.00 (0.00)*** & \phantom{-}0.00 (0.00)    \\ 
  \% black offenders (ORI) & -0.20 (0.16)    & -0.43 (0.05)*** & -0.66 (0.05)*** & -0.99 (0.02)*** & -0.97 (0.01)*** \\ 
  Population served (ORI) & \phantom{-}0.00 (0.00)*** & \phantom{-}0.00 (0.00)*   & \phantom{-}0.00 (0.00)*** & \phantom{-}0.00 (0.00)*** & \phantom{-}0.00 (0.00)*** \\ 
  Core city & \phantom{-}0.26 (0.07)*** & -0.21 (0.02)*** & -0.12 (0.02)*** & -0.02 (0.01)*   & \phantom{-}0.04 (0.00)*** \\ 
   \bottomrule
\end{tabular}}
  
  {\raggedright {\footnotesize Significance codes: p$-$value<0.001 ‘***’,
  $p<0.01$ ‘**’, $p<0.05$ ‘*’, $p<0.1$ ‘.’\\
  Notes: As a reminder, we only consider incidents involving lone victims and
  offenders of race Black or White (Hispanics are included), that occurred in the 16 states considered in the
  analysis between 2007 and 2016, and were not
  cleared by exceptional means. Coefficients estimates of dummies for states and years are omitted from the table. 
  For a discussion of the estimates, see
  \textsection\ref{sec:reg_analysis}.}\par}
\end{table*}

\end{document}